\documentclass[twocolumn,prl,aps,floatfix]{revtex4}
\usepackage{amsmath,amssymb}
\usepackage{graphicx,graphics}
\usepackage{fancyhdr}

\begin{document}

\unitlength=1mm

\title{Excitations of strongly correlated lattice polaritons}
\author{S.\ Schmidt}
\author{G.\ Blatter}
\affiliation{Institute for Theoretical Physics, ETH Zurich, 8093 Zurich, Switzerland}

\date{\today}

\begin{abstract}

We present an analytic slave-boson approach to calculate the elementary excitations of the Jaynes-Cummings-Hubbard Model (JCHM)
describing strongly correlated polaritons on a lattice in various quantum optical systems. 
In the superfluid phase near the Mott transition we find a gapless, linear Goldstone mode and a gapped 
Amplitude mode corresponding to phase and density fluctuations, respectively. 
The sound velocity of the Goldstone mode develops a peculiar anomaly as a function of detuning at low densities, 
which persists into the weakly interacting regime of a polariton BEC.

\end{abstract}


\maketitle
The realization of a BEC of weakly interacting polaritons in high-Q microcavities \cite{KR06}, 
i.e., quasiparticles which form when photons strongly interact with quantum well excitons, 
has triggered an immense interest in quantum condensed and coherent light-matter systems.
The addressibility of a cavity with external lasers, device integration, and high tunability make these optical 
systems ideally suited for quantum information and simulation tasks. Furthermore, the inherently composite nature 
of polaritons promises rich and interesting new physics. 

One of the most exciting questions in this emerging field is whether one can realize a Mott insulator - superfluid (MI-SF) transition of strongly correlated polaritons. 
The Jaynes-Cummings-Hubbard Model (JCHM) has been introduced to describe such a quantum phase transition of light
in an array of coupled QED cavities, each containing a single photonic mode interacting with a two-level system (qubit) \cite{GT06}.
Meanwhile the JCHM has become a paradigm lattice model for polaritons describing various systems at the interface of condensed matter
and quantum optics \cite{HB08}. Proposals based on cavity QED arrays can be realized using atoms \cite{Bi05},  excitons \cite{HB07}, or superconducting qubits \cite{FG07}.
In trapped ion systems, e.g., in a linear Paul trap, a transverse phonon assumes the role of the photon and couples to hyperfine levels
of the ion via an external laser \cite{II09}; tunneling of phonons is mediated by Coulomb interactions between the ions.
Ultracold atom systems with two optical lattices that couple to the photon mode of a single QED cavity may also realize a JCHM \cite{BH09}.

Theoretically, the phase diagram and the excitations in the Mott phase have been calculated accurately in $D=1,2$, and $3$ dimensions \cite{RF07,ZS08,AH08,SB09,KH09,PE09,KA09}.
However, very little is known about the nature of the superfluid phase, except the existence of a Goldstone mode predicted by heavy Monte-Carlo simulations in $D=1$ \cite{PE09}.
In this letter we present an analytic slave-boson approach valid in both phases.
In the Mott phase we find gapped particle and hole modes in quantitative agreement with previous results obtained
from a diagrammatic linked-cluster expansion \cite{SB09}. In the superfluid phase we find a gapless linear Goldstone mode and a 
gapped Amplitude mode corresponding to phase and density fluctuations (similar excitations were found in ultracold gases \cite{AA02}). 

In addition, our analytic approach is ideally suited to discuss the behavior of excitations in the superfluid phase as a function of detuning, which
provides an important experimentally accessible parameter. Detuning can be used to change (i) the nature of polaritonic excitations, i.e., the relative weight of
its bosonic and atomic parts, and (ii) the strength of the effective repulsion between polaritons.
We find that the phase diagram as well as the sound velocity of the Goldstone mode show a surprising anomaly when this parameter is varied:
For high polariton densities, the size of the Mott lobes and the sound velocity decrease for any finite detuning, while at low densities 
they steadily increase when tuning through the resonance. 
Comparison to an effective Bogoliubov theory for the JCHM shows that this anomaly 
prevails into the regime of a weakly interacting polariton BEC \cite{KE04}. In this regime, superfluidity has been demonstrated recently by measuring 
the sound velocity of exciton-polaritons in a single micro-cavity using angle-resolved photon spectroscopy \cite{UT08}.  
Similar experimental techniques should also be applicable to lattice models as discussed here.

The Hamiltonian of the JCHM is given by
\begin{equation}
\label{jchm0}
H=\sum_i h^{\rm JC}_i - \mu N - J \sum_{\langle i j \rangle} a^\dagger_i a_j\,,
\vspace{-0.1cm}
\end{equation}
where $h^{\rm JC}_i$ denotes the local Jaynes-Cummings Hamiltonian 
$h^{\rm JC}_i = \omega_c\, a^\dagger_i a_i + \omega_x \sigma_{i}^+\sigma_{i}^-  +  g (\sigma_{i}^+ a_i +\sigma_{i}^- a^\dagger_i)$
with site index $i$, boson creation (annihilation) operators $a_i^{(\dagger)}$ and qubit raising (lowering) operators $\sigma_i^{+(-)}$. 
The bosonic mode frequency is $\omega_c$, the two qubit levels are separated by the energy $\omega_x$ and the coupling is given by $g$ (we set $\hbar=1$).  We also assume that the total number of excitations, i.e., polaritons $N=\sum_i ( a^\dagger_i a_i + \sigma_{i}^+\sigma_{i}^-)$, is conserved and fixed by the chemical potential $\mu$.  We note that even under more general conditions including the effects of an external drive and dissipation the underlying equilibrium model still captures the essential physics of the MI-SF transition as recently shown for the Bose-Hubbard model (BHM) \cite{TG09}.
The third term in (\ref{jchm0}) describes the delocalization of bosons over the whole lattice due to hopping between nearest neighbour sites with amplitude $J$. It competes with an effective on-site repulsion between bosons mediated by the coupling $g$. This competition leads to Mott lobes in the quantum phase diagram  \cite{GT06}.

The on-site eigenstates of the Jaynes-Cummings Hamiltonian $h^{\rm JC}_i$ are labelled by the polariton number $n$ and upper/lower branch index $\sigma=\pm$. 
The mixed boson ($n,n-1$) - qubit ($g,e$) states define upper and lower polariton states
\begin{eqnarray}
|n +\rangle &=& \sin\theta_n |n\,,g\rangle + \cos\theta_n |(n-1)\,,e\rangle\,, \nonumber\\
|n -\rangle &=& \cos\theta_n |n\,,g\rangle -  \sin\theta_n |(n-1)\,,e\rangle\,,
\end{eqnarray}
with the mixing angle $\tan \theta_n=2 g \sqrt{n}/(\delta+2\chi_n)$, $\chi_n=\sqrt{g^2 n + \delta^2/4}$ and the detuning parameter $\delta=\omega_c-\omega_x$.
The corresponding eigenvalues are
\begin{equation}
\epsilon_n^{\sigma}=-(\mu-\omega_c) n - \delta/2 +\sigma\,\chi_n\,,\quad \sigma=\pm\,.
\end{equation}
The zero polariton state $|0\rangle\equiv|0-\rangle=|0\,,g\rangle$ is a special case with $\epsilon_0\equiv\epsilon_0^-=0$. 

A convenient starting point for our slave-boson approach is the polariton representation \cite{KH09} of the boson operator $a_i=\sum_{n \sigma \nu} f^{\sigma \nu}_n P^{\nu\dagger}_{i n-1} P^\sigma_{i n}$ in terms of standard algebra operators $P^{\sigma\dagger}_{i n}=|n \sigma\rangle_{ii}\langle 0 |$ and matrix elements $f^{\sigma \nu}_n= \langle n-1\,\nu | a | n\, \sigma\rangle$
with $f^{\sigma \nu}_n=\left( \sqrt{n}+\sigma\,\nu\,\sqrt{n-1}\right)/2$ for $n>1$  ($f^{\sigma -}_1=1/\sqrt{2}$) at zero detuning ($\delta=0$). 
In this new basis the JCHM becomes
\begin{eqnarray}
\label{jchm}
H&=&\sum_i\sum_{n=0}^\infty\sum_\sigma \epsilon^\sigma_{n} {P^{\sigma\dagger}_{i n}} P^\sigma_{in}\\ \nonumber
&-&J\sum_{\langle ij\rangle}\sum_{n,n'=1}\mathop{\sum_{\scriptsize\sigma,\sigma'}}_{\nu,\nu'}f^{\sigma\sigma'}_n f^{\nu\nu'}_{n'}\, P^{\sigma\dagger}_{i n} P^{\sigma'}_{i n-1} P^{\nu' \dagger}_{j n'-1} P^\nu_{j n'}\,.
\end{eqnarray}
The polariton operators obey bosonic commutation relations if the constraint 
\begin{equation}
\label{constraint}
\sum_{n\sigma} P^{\sigma\dagger}_{i n} P^\sigma_{in}=1
\end{equation}
is fullfilled at each site $i$. 
\begin{figure}
\includegraphics[scale=0.58]{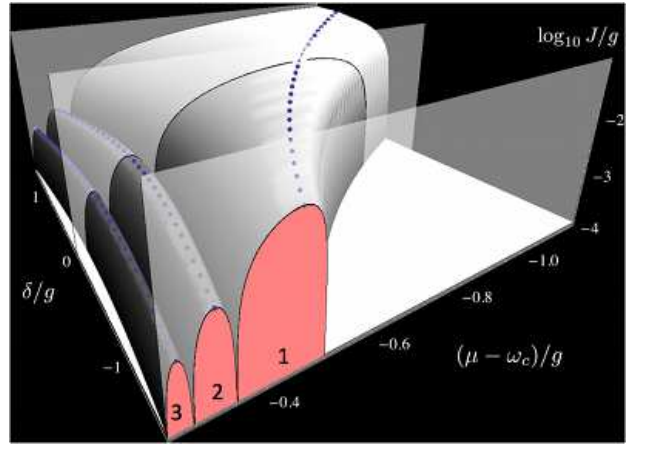}
\vspace{-0.2cm}
\caption{\label{fig1} Phase diagram for the JCHM displaying the lowest three Mott lobes with $n=1,2,3$. Dotted lines represent the 
critical hopping strength's $J_c/g$, where chemical potential and detuning are chosen such as to fullfill particle-hole symmetry.
A finite detuning $|\delta|>0$ decreases the critical hopping strength $J_c/g$ for $n>1$, but the lowest Mott lobe ($n=1$) steadily increases when
tuning through the resonance ($\delta=0$).}
\vspace{-0.3cm}
\end{figure}
In \cite{SB09} we have shown that the presence of the upper polariton branch ($\sigma=+$) leads to additional high energy
conversion modes in the Mott phase with small spectral weight and bandwidth; we neglect the upper 
branch as well as particle conversion tunneling from now on and drop the branch index $\sigma$. 
In order to calculate the phase boundary and static observables in the superfluid phase near
a Mott lobe with filling $n$, we restrict the Hilbert space to states with $n$ and $n\pm1$ bosons
and make a Gutzwiller Ansatz for the ground-state wave function
\begin{equation}
|\psi\rangle=\hspace{-0.1cm}\prod_i\hspace{-0.1cm}\big[ \cos(\theta) P^\dagger_{i 0} \hspace{-0.05cm}+ \sin(\theta)( \sin(\chi) P^\dagger_{i -1} \hspace{-0.06cm}+\hspace{-0.05cm} \cos(\chi) P^\dagger_{i 1}) \big]|0\rangle\nonumber
\end{equation}
where we also dropped the index $n$, i.e., $P^\dagger_{i\alpha}\equiv P^\dagger_{in+\alpha}$, $\epsilon^\sigma_{n+\alpha}\equiv \epsilon_\alpha$, and $f^{\sigma\nu}_{n+\alpha}\equiv f_\alpha$.
This yields the variational energy $\epsilon_{\rm var}=\langle \psi | H | \psi \rangle$, 
\begin{eqnarray}
\label{evar}
\epsilon_{\rm var}&=&\epsilon_0 \cos(\theta)^2+\sin(\theta)^2\left( \epsilon_{-1}\sin(\chi)^2+\epsilon_{1}\cos(\chi)^2\right)\nonumber\\
&-& J z/4 \sin(2\theta)^2\left( f_0 \cos(\chi) + f_{-1} \sin(\chi)\right)^2\,,
\end{eqnarray}
which has to be minimized with respect to the variational parameters $\theta$ and $\chi$. The lobe boundaries as shown in Fig.~\ref{fig1} are determined by
the vanishing of the order parameter $\phi_c=\langle \psi | a | \psi\rangle=\sin(\theta)\left[f_0\sin(\chi)+f_{1}\cos(\chi)\right]^2/2$.
Our results for the quantum phase diagram agree exactly with those in \cite{SB09,KH09}.
Here, we point out that the size of all Mott lobes with filling factor $n>1$ decrease for any finite detuning $|\delta|>0$, while the size of the lowest Mott lobe ($n=1$)
increases steadily as the system is tuned through the resonance ($\delta=0$). We explain this special behavior when discussing the nature of excitations close to the lobes below.

In order to find the excitations we follow the procedure outlined
in \cite{AA02} for the Bose-Hubbard model and define a new set of operators ${\bf R}^\dagger=(G_i^\dagger, E_{1i}^\dagger, E_{2i}^\dagger)^T$, which is 
obtained from the original polariton basis ${\bf P}^\dagger=(P_{i 0}^\dagger, P_{i -1}^\dagger, P_{i 1}^\dagger)^T$
via a unitary transformation ${\bf R}^\dagger=T {\bf P}^\dagger$ with
\begin{eqnarray}
T=\left( \begin{array}{ccc} \cos(\theta) & \sin(\theta)\cos(\chi) & \sin(\theta)\sin(\chi) \\ 
-\sin(\theta) & \cos(\theta)\cos(\chi) & \cos(\theta)\sin(\chi) \\ 0 & -\sin(\chi) & \cos(\chi) \end{array} \right)\,.
\end{eqnarray}
The operator $G^\dagger$ creates a new vacuum state, i.e., the mean-field ground state $|\psi\rangle=\prod_i G_i^\dagger |0\rangle$,
and $E_{1i}^\dagger, E_{2i}^\dagger$ are orthogonal operators creating excitations above the ground-state.
We express the Hamiltonian in terms of these new operators and eliminate $G_i$ by using the constraint (\ref{constraint})
in the restricted Hilbert space $G_i\approx\sqrt{1-E_{1i}^\dagger E_{1i}-E_{2i}^\dagger E_{2i}}$.
Expanding the square root  everywhere in the Hamiltonian to quadratic order in $E^{(\dagger)}_{(1,2)i}$ yields,
after a Fourier transformation, an effective quadratic Hamiltonian
$H_{\rm eff}=\epsilon_{\rm var}+\sum_{\bf k} {\bf E}_{\bf k}^\dagger\, h_{{\rm eff},{\bf k}} \,{\bf E}_{\bf k}$,
where ${\bf E}=(E_{1{\bf k}},E_{2{\bf k}},E_{1-{\bf k}}^\dagger,E_{2-{\bf k}}^\dagger)^T$
and $h_{{\rm eff},{\bf k}}$ is a $4\times 4$ matrix. The sum over ${\bf k}$ runs over the first Brioullin zone.
The effective Hamiltonian can be diagonalized by a bosonic Bogoliubov transformation \cite{WS65} yielding
\begin{equation}
H_{\rm eff}=\epsilon_{\rm var}+\epsilon_{\rm fluct}+\sum_{\alpha=1,2}\sum_{\bf k} \epsilon_\alpha({\bf k}) d_{\alpha{\bf k}}^\dagger d_{\alpha{\bf k}}
\end{equation}
with $\epsilon_{\rm fluct}$ a fluctuation-generated correction of the ground-state energy
and $d_{\alpha {\bf k}}^\dagger$ creating excitations with energy $\epsilon_{\alpha}({\bf k})$.
\begin{figure}[t]
\includegraphics[scale=0.6]{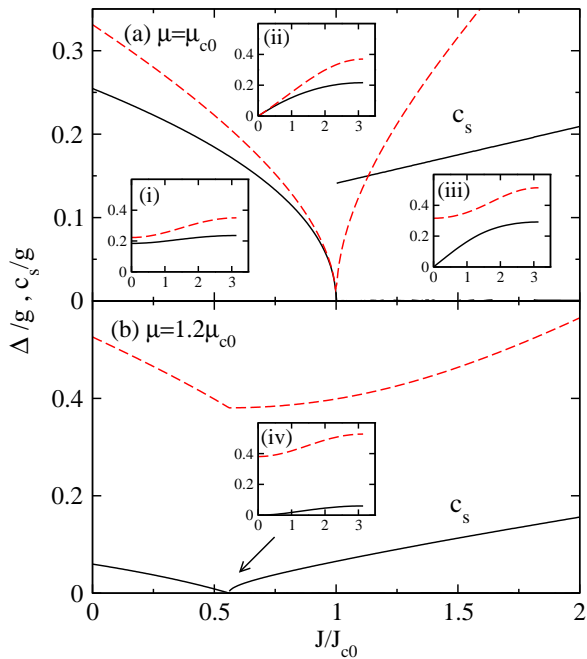}
\vspace{-0.2cm}
\caption{\label{fig2} Spectral gaps and sound velocity as a function of tunneling strength $J/J_{c0}$ at zero detuning $\delta/g=0$ and for 
(a) $\mu=\mu_{c0}$ where $\mu_{c0}$ denotes the critical chemical potential at the tip of the lobe with critical hopping strength $J_c=J_{c0}$ (top figure) 
and (b) away from the tip at $\mu=1.2\mu_{c0}$ with  $J_c=0.566J_{c0}$ (bottom figure).
Shown are the gaps of particle (dashed) and hole (solid) modes in the Mott phase ($J<J_c$) and the gaps 
of the Amplitude mode (dashed) and the sound velocity of the Goldstone mode (solid) in the superfluid phase ($J>J_{c}$). The insets show the corresponding excitation spectra at i) $J=0.5J_{c0}$ 
ii) $J=J_{c0}$ iii) $J=1.5J_{c0}$ iv) $J=0.566J_{c0}$.
At the phase boundary, the particle and hole mode of the Mott phase are identical with the Goldstone and Amplitude modes of the superfluid phase. 
At the tip of the lobe ii), where the polariton density can remain constant during the superfluid-insulator transition, the Amplitude mode becomes gapless and linear (its mass vanishes).
The sound velocity of the Goldstone mode remains non-zero, confirming a special point in the phase diagram with dynamical critical exponent $z=1$.
Away from the tip iv), the Amplitude mode remains gapped and the Goldstone mode becomes quadratic with a vanishing sound velocity corresponding
to a generic dynamical critical exponent $z=2$. This confirms the change of the universality class along the phase boundary found in \cite{SB09,KH09}, but put into
question in \cite{ZS08}.}
\vspace{-0.5cm}
\end{figure}
The analytic expressions for the two modes $\epsilon_{\alpha}({\bf k})$ are rather lengthy and will be presented
elsewhere. They depend on the variational parameters $\theta$ and $\chi$. In order to calculate the excitations at different points 
in the phase diagram, we first determine the optimal parameters $\theta_{\rm opt}$ and $\chi_{\rm opt}$ from a minimization of (\ref{evar}) and insert them into the corresponding algebraic 
expressions for the spectra $\epsilon_{\alpha}({\bf k})$. The results are summarized in Fig.~\ref{fig2}. In the Mott phase, where $\theta_{\rm opt}=0$, we find two gapped modes corresponding to particle and hole excitations of lower polaritons. Our results agree exactly with the ones obtained from a strong-coupling RPA \cite{SB09}.  In the superfluid phase we obtain a gapless Goldstone mode which is linear for small ${\bf k}$ describing the propagation of
phonons with a characteristic velocity $c_s$ and a  gapped Amplitude mode, which describes local density fluctuations between normal fluid and condensate. Its existence has been predicted for cold atoms in an optical lattice as described by the BHM \cite{AA02}.

\begin{figure}
\includegraphics[scale=0.58]{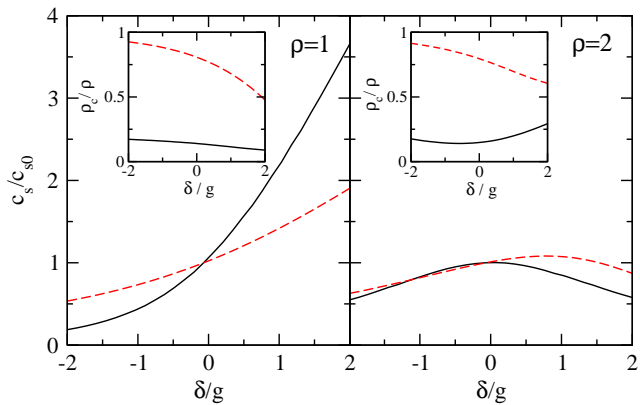}
\vspace{-0.4cm}
\caption{\label{fig3} Sound velocity as a function of detuning $\delta/g$ for fixed polariton density $\rho=1$ (left figure) and $\rho=2$ (right figure) near the Mott lobes with $J=J_{c0}(\delta)+0.2 J_{c0}(0)$ (solid lines) obtained from slave-boson theory and deep inside the superfluid phase with $J=J_{c0}(\delta)+5 J_{c0}(0)$ (dashed lines) obtained from an effective Bogoliubov theory. Here, $J_{c0}(\delta)$ denotes the critical hopping strength at the tip of the lobe. The insets show the condensate fraction $\rho_c/\rho$ with the same notation as above.}
\vspace{-0.5cm}
\end{figure}

In cavity QED systems temporal and spatial correlation functions are experimentally accessible and provide an important tool
to characterize the nature of different phases. 
In particular, the sound mode is a key signature of superfluidity and an important quantity which signals the presence of interactions in a condensate. 
In Fig.~\ref{fig3}, we study the dependence of the sound velocity $c_s$ and the condensate density $\rho_c=|\phi_c|^2$ on detuning $\delta/g$.

In the strong-coupling regime ($J\approx J_c$) the sound velocity develops an anomaly vs. detuning at low polariton densities $\rho\equiv\langle N \rangle/N_s$ (with $N_s$ the number of lattice sites).
While for $\rho=2$ (and higher) $c_s$ decreases for any finite detuning $|\delta|>0$, it increases steadily for $\rho=1$ when 
tuning through the resonance.
To explain this behavior, we define the particle-hole gap in the atomic limit $U_{n}(\delta)=\epsilon_{n+1}-2\epsilon_n+\epsilon_{n-1}$. This provides a 
useful measure for the effective repulsive interaction between polaritons and its dependence on detuning near the Mott lobes, where the mixing of 
states with different polariton numbers is small. For the Bose-Hubbard model $U_{n}(\delta)=U$ is simply the Hubbard interaction parameter but for the JCHM the particle-hole gap depends
on filling $n$ and detuning $\delta/g$. For $n \gg 1 $ we find $U_{n}(\delta)-U_{n}(0)=-(3/8n^{5/2})(\delta^2/g)+\mathcal{O}(\delta^4/g^3,1/n^{7/2})$ 
and thus $U_{n}$ decreases for any finite detuning $|\delta|>0$, consistent with a decrease of the sound velocity.
However, for $n=1$ we find  $U_{1}(\delta)-U_{1}(0)=\delta/2+\mathcal{O}(\delta^2/g)$, i.e., the effective repulsion increases steadily when tuning through the resonance. This causes the
anomaly in $c_s$ at low densities. The different behavior of $U_{1}$  is rooted in the special nature of
the zero-polariton state with $\epsilon_0=0$ independent of detuning. 
This special behavior is also the reason for the decrease of the condensate fraction $\rho_c/\rho$ for $\rho=1$ (see inset in Fig.~\ref{fig3}) and the increase of the size of the
lowest Mott lobe (see phase diagram in Fig.~\ref{fig1}). At higher densities the condensate fraction increases while the size of
the lobes decrease for any finite detuning, consistent with a decrease of the effective repulsion.
We conclude that in the strongly-correlated regime both sound velocity $c_s$ and condensate fraction $\rho_c/\rho$ are dominated by the repulsive interaction between polaritons
when $\delta/g$ is varied.\\
In the weakly interacting regime ($J\gg J_c$), the slave-boson approach is no longer a good approximation since states with different polariton numbers
mix strongly. We use an effective Bogoliubov theory instead, originally derived for a generalized Dicke model describing localized excitons 
coupled to photons confined in a single microcavity with a gapped, quadratic photon dispersion $\epsilon(k)=\omega_0+{\bf k}^2/2m$ with $\omega_0=\pi c/L$ and $m=\pi /(c L)$
($L$ being the width of the cavity) \cite{KE04}. The JCHM maps onto the generalized Dicke model after a Fourier transformation of the bosonic field operator and expanding the 
lattice dispersion for small ${\bf k}$-vectors as $\epsilon(k)\approx J z+J {\bf k}^2$ (for a hypercubic lattice with lattice constant $a=1$). 
Using the results in \cite{KE04} with the formal replacements $\omega_0=\omega_c-zJ$ and $m=1/(2J)$, we can compare our findings in the strongly interacting regime with those valid for weak interactions shown as
dashed lines in Fig.~\ref{fig3}. We observe that the sound anomaly is still present deep inside the superfluid phase, but slightly suppressed. 
This is due to the admixture of states with higher polariton numbers $n>1$ for $\rho=1$. 
On the other hand, the maximum of the sound velocity at higher densities $\rho=2$ is slightly shifted to positive detuning due to a weak admixture of polariton states with $n=0,1$.
On the contrary, the condensate density behaves different in the weakly interacting regime far away from the Mott lobes. 
It decreases independent of the polariton density when tuning through the resonance and is thus no longer dominated by the effective 
repulsion but rather by the internal structure of the polaritons. 
To understand this we have to consider the boson density $\rho_{\rm B}=\rho_n+\rho_c$, which consists of a condensate $\rho_c$ and a normal part $\rho_n$ (depletion).  
In the weakly interacting regime, the depletion $\rho_n$ is small and the condensate density $\rho_c$ behaves similar to the density of bosonic excitations $\rho_{\rm B}$; 
both decrease when the energy cost of creating a boson is increased with respect to the energy cost of an atomic excitation, i.e., when tuning through the resonance.

In summary, the slave-boson theory for strongly correlated polaritons presented in this paper predicts the existence of gapped particle/hole modes in the
Mott regime and, besides a gapless, linear Goldstone mode, the existence of a gapped Amplitude mode in the superfluid phase.
The anomaly of the sound velocity of the Goldstone mode as a function of detuning and polariton density is rooted in the special nature of the zero polariton state within the 
Jaynes-Cummings ladder. It is present in the strong as well as weak coupling regime and may constitute an interesting experimental signature for the composite 
nature of superfluid lattice polaritons in future experiments with coupled cavities.
\begin{acknowledgments}
We thank H. Tureci and S. Huber for discussions and acknowledge financial support from the Swiss National Foundation through the NCCR MaNEP. 
\end{acknowledgments}

\vfill
\end{document}